\documentclass[twocolumn,superscriptaddress,aps,preprintnumbers,amsmath,amssymb,prl,nofootinbib]{revtex4}
\usepackage{graphicx}
\usepackage{epstopdf}
\usepackage{dcolumn}
\usepackage{bm}
\usepackage{hyperref}
\usepackage{color}

\def\b{\beta}

\def\l{\lambda}

\def\D{\Delta}

\def\L{\Lambda}

\def\beq{\begin{eqnarray}}
\def\eeq{\end{eqnarray}}

\begin{document}
\

\title{
Flat Higgs Potential from Planck Scale Supersymmetry Breaking
}

\author{Masahiro Ibe}
\affiliation{IPMU, University of Tokyo, Kashiwa, 277-8568, Japan}
\affiliation{ICRR, University of Tokyo, Kashiwa, 277-8582, Japan}
\author{Shigeki Matsumoto}
\affiliation{IPMU, University of Tokyo, Kashiwa, 277-8568, Japan}
\author{Tsutomu T.~Yanagida}
\affiliation{IPMU, University of Tokyo, Kashiwa, 277-8568, Japan}

\begin{abstract}
The observed Higgs boson mass poses a new puzzle in addition to the longstanding problem
of the origin of the electroweak scale; the shallowness of the Higgs potential.
The Higgs quartic coupling even seems to vanish at around the Planck scale within 
the uncertainties of the top quark mass and the strong gauge coupling.
We show that the shallowness of the Higgs potential might be an outcome
of supersymmetry breaking at around the Planck scale.
There, the electroweak fine-tuning in the Higgs quadratic terms 
leads to an almost vanishing quartic coupling at around the Planck scale.
\end{abstract}

\date{\today}
\maketitle
\preprint{IPMU13-0245}
With the discovery of the Higgs boson at the LHC experiments\,\cite{Aad:2012tfa,Chatrchyan:2012ufa},
the investigation of the detailed structure of the Higgs sector has just started.
Among other things, the measured Higgs boson mass, $m_h = 125.9\pm 0.4$\,GeV\,\cite{Beringer:1900zz},
seems to pose a new puzzle in addition to the longstanding problem
of the origin of the electroweak scale; why the Higgs potential is so shallow.
In fact, the extrapolated Higgs quartic coupling seems to vanish at around
the Planck scale within the uncertainties of the top quark mass and the 
strong gauge coupling if we assume that there are no new physics 
below the Planck scale\,\cite{Holthausen:2011aa,EliasMiro:2011aa,Degrassi:2012ry,Buttazzo:2013uya}.

So far, a lot of attempts to provide such a boundary condition of the flat Higgs potential
at around the Planck scale have been discussed based on,
such as the asymptotic safety\,\cite{Shaposhnikov:2009pv}, 
or the multiple point criticality principle\,\cite{Froggatt:1995rt}
(for recent works, see e.g.\,Ref.\cite{Iso:2012jn,Hashimoto:2013hta,Holthausen:2013ota,Haba:2013lga}).  
In this letter, we propose a new possibility where the almost vanishing quartic Higgs coupling at the Planck scale
is an outcome of  supersymmetry breaking at around the Planck scale.
As we will show, the electroweak fine-tuning in the Higgs mass parameters 
automatically leads to an almost vanishing quartic coupling
either when the supersymmetry breaking sector is weakly coupled to the Higgs sector,
or when the soft squared masses of the two Higgs doublets are close with each other.

\subsubsection*{Fine-tuning in the Higgs quadratic terms}\vspace{-10pt}
To explain how the quartic coupling constant is determined at around the Planck scale, $M_{\rm PL}$,
let us take the simplest Higgs sector as an example, where  
the K\"ahler and the superpotential are given  by
\begin{eqnarray}
\label{eq:Kp}
K &=& Z^\dagger Z + H_u^{\dagger} H_u + H_d H_d^\dagger + (c H_u H_d + h.c.)\ , \\
\label{eq:Sp}
W & = &  \Lambda_{\rm SUSY}^2 Z + m_{3/2}M_{\rm PL}^2\ .
\end{eqnarray}
Here, $c$ denotes a dimensionless constant of $O(1)$, and 
$\L_{\rm SUSY}$ and $m_{3/2}$ are the supersymmetry breaking scale and the gravitino mass, respectively.
The supersymmetry breaking field $Z$ obtains an $F$-term vacuum expectation value, $F_z = - \L_{\rm SUSY}^2$,
and the flat universe condition gives $\L_{\rm SUSY}^4 \simeq 3 m_{3/2}^2 M_{PL}^2$.
We assume that the supersymmetry breaking scale is at around the Planck scale 
and higher dimensional operators which couple supersymmetry breaking field 
and the Higgs doublets are somehow suppressed.

With these potentials, the Higgs mass terms are given by,
\begin{eqnarray}
\label{eq:mass}
V_2 &=& \bar{m}_{H_u}^2 |H_u|^2 
+ \bar{m}_{H_d}^2 |H_d|^2 + (b H_u H_d + h.c.) \nonumber \\
 &\simeq& (|\mu_H|^2 + m_{3/2}^2) |H_u|^2 
+ (|\mu_H|^2 + m_{3/2}^2) |H_d|^2  \nonumber \\
 & &+ (b H_u H_d + h.c.)\ ,
\end{eqnarray}
where  $\mu_H$ and $b$ are given by,
\begin{eqnarray}
\mu_H = c m_{3/2} \ , \quad
b  =  2 cm_{3/2}^2\ .
\end{eqnarray}
Hereafter, we take $b$ to be real  and positive by  redefining the phases of $H_u$ and $H_d$ appropriately.

The higher dimensional operators which couple the Higgs doublets to the supersymmetry 
breaking field such as,
\begin{eqnarray}
K = \frac{c_{u,d}}{M_{\rm PL}^2} |Z|^2 |H_{u,d}|^2 \ ,
\end{eqnarray}
lead to additional contributions to the Higgs mass parameters.
In the followings, we assume that the coefficients are rather suppressed, i.e. $c_{u,d}<{\cal O}(0.1)$
or the coefficients are almost universal,
 i.e. $(c_u - c_d)/(c_u + c_d) < {\cal O}(0.1)$,
so that the soft squared masses of the two Higgs doublets are close with each other, i.e. $\bar{m}_{H_u}^2 \simeq \bar{m}_{H_d}^2$.%
\footnote{
For example, the almost universality can also be realized by an approximate symmetry which interchanges $H_u$ and $H_d$.
}

For  successful electroweak symmetry breaking,
we need fine-tuning so that one of the linear combinations of the two Higgs doublets, $h = \sin\b H_u -\cos\b H_d^\dagger$,
remains very light with a mass much smaller than the Planck scale.
In terms of the  Higgs mass parameters, this requires
\begin{eqnarray}
\label{eq:FT}
\bar{m}_{H_u}^2 \bar{m}_{H_d}^2 - b^2 \ll {\cal O}( M_{\rm PL}^4 )\ ,
\end{eqnarray}
which leads to $b \simeq \bar{m}_{H_u}^2 \simeq \bar{m}_{H_d}^2$.
Therefore, by remembering that the Higgs mixing angle is determined by
\begin{eqnarray}
\tan2\b \simeq \frac{2 b}{\bar{m}_{H_u}^2 - \bar{m}_{H_d}^2 }\ ,
\end{eqnarray}
we find that the electroweak fine-tuning predicts 
\begin{eqnarray}
|\tan2\b| \gg 1 \ ,
\end{eqnarray}
and hence,
\begin{eqnarray}
\tan\b \simeq 1 \  ,
\end{eqnarray}
for almost universal Higgs doublet masses, $\bar{m}_{H_u}^2 \simeq \bar{m}_{H_d}^2$.

At  higher loop levels, the radiative corrections change the mass parameters in  Eq.\,(\ref{eq:mass}).
Thus, the fine-tuning condition and the Higgs mixing angle are accordingly changed to
\begin{eqnarray}
\label{eq:FTH}
(\bar{m}_{H_u}^2+{\Delta}_{H_u}) (\bar{m}_{H_d}^2+ {\Delta}_{H_d}) - (b^2+\Delta_b)\cr \ll {\cal O}( M_{\rm PL}^4 )\ ,
\end{eqnarray}
\begin{eqnarray}
\tan2\b \simeq \frac{2 (b+ \D_b)}{(\bar{m}_{H_u}^2 + \D_{H_u}) -( \bar{m}_{H_d}^2 + \D_{H_d})}\ .
\end{eqnarray}
Here, $\D_{H_u,H_d,b}$ denote the radiative corrections to the mass parameters.%
\footnote{In terms of the K\"ahler potential, the radiative corrections to the Higgs parameters leads to
\begin{eqnarray}
K \simeq (1 + \delta_{u,d} |Z|^2/M_{\rm PL}^2)|H_{u,d}|^2 + ( (c + \delta_c) H_u H_d + h.c.) \ .
\end{eqnarray}
where $\delta$'s are expected to be small.
 }
The fine-tuning condition for the light Higgs boson 
is imposed only after  the radiative corrections to the mass parameters of all orders are included.
Those corrections are, however, expected to be at most about a 10\% compared to the tree-level mass parameters
since the Standard Model interactions are rather suppressed 
at around the Planck scale (e.g. the top Yukawa coupling is $y_t \simeq 0.4$ at around the Planck scale).
Therefore, the prediction of $\tan\b \simeq 1$ at the tree level is not significantly affected by the radiative 
corrections.

\begin{figure}[t]
\begin{center}
  \includegraphics[width=0.75\linewidth]{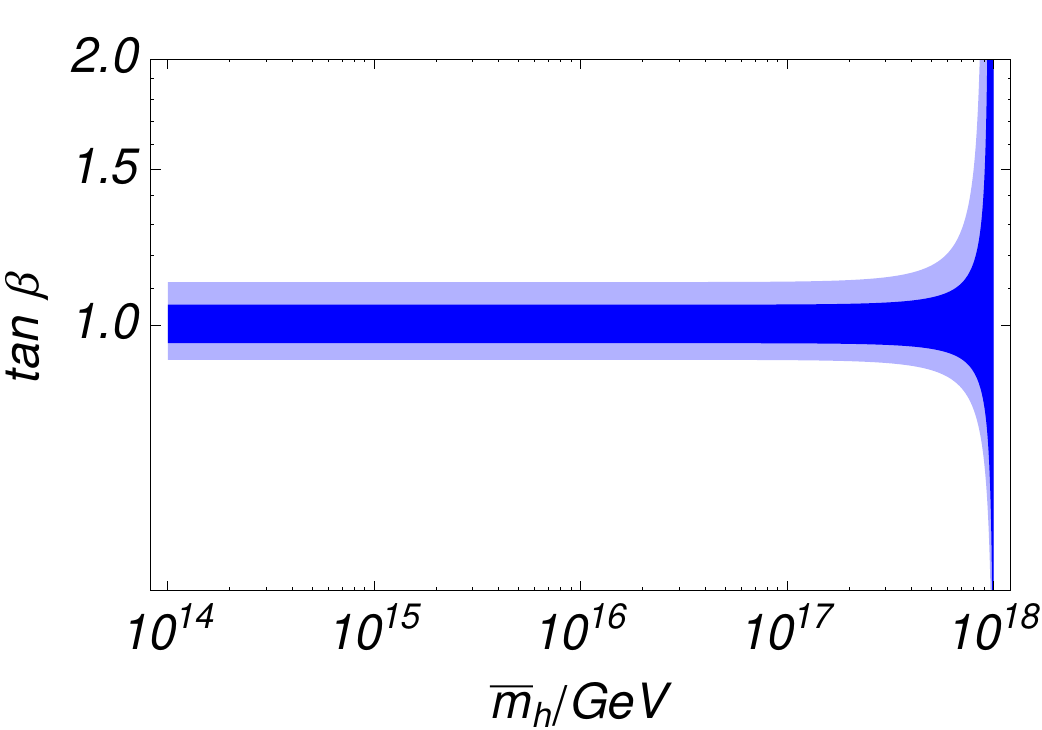}
\end{center}
\caption{\sl \small
The predicted value of $\tan\b$ as a function
of finely tuned light Higgs boson mass parameter $\bar{m}_h$.
We varied $\bar{m}_{H_u}^2$ from $\bar{m}_{H_d}^2$ 
by $10$\% (blue band) and $20$\% (light blue band).
}
\label{fig:tanb}
\end{figure}

In Fig.\,\ref{fig:tanb}, we show the predicted value of $\tan\b$ as a function the fine-tuned mass parameter of the light Higgs boson;
\begin{eqnarray}
\bar{m}_h^2 &\simeq& \frac{1}{2}\left\{ (\bar{m}_{H_u}^2 + \bar{m}_{H_d}^2)\right. \nonumber\\ 
&&\left.- \sqrt{\bar{m}_{H_u}^4 + \bar{m}_{H_d}^4- 2 \bar{m}_{H_u}^2 \bar{m}_{H_d}^2  + 4 b^2 } 
\right\} \ .
\end{eqnarray}
Here, we redefined the mass parameters in the right hand side so that they include the radiative corrections.
In the figure, we varied $\bar{m}_{H_u}^2$ from $\bar{m}_{H_d}^2$ 
by $10$\% (blue band) and $20$\% (light blue band) to explore how the non-universality as well as 
the  radiative corrections change the prediction.
The figure shows that for $\bar{m}_h \ll M_{\rm PL}$, the predicted value of $\tan\b$
immediately converges to $\tan\b \simeq 1$.
The figure also shows that the prediction is not significantly affected even when $\bar{m}_{H_u}^2$
deviates from $\bar{m}_{H_d}^2$ by $20$\%.

It should be noted that unlike the low energy supersymmetry,
the renormalization group effects to the Higgs mass parameters are negligible.
For example, the up-type Higgs  squared mass receives a correction 
from the renormalization group effects,
\begin{eqnarray}
{\mit \Delta}m_{H_u}^2 \simeq \frac{6y_t^2}{16\pi^2} m_{\tilde t}^2 \log \frac{m_{3/2}^2}{M_{\rm PL}^2}\ ,
\end{eqnarray}
where  $m_{\tilde t}$ denotes the typical mass of the top squarks. 
These corrections are, however, not significant and lead to only a few percent changes to the Higgs mass parameters, $\bar{m}_{H_{u,d}}^2$, 
as long as $m_{3/2} \simeq M_{\rm PL}$ and $m_{\tilde t} \simeq m_{3/2}$.

Let us emphasize that the prediction of $\tan\b \simeq 1$ is not altered 
as long as $\bar{m}_{H_u}^2 \simeq \bar{m}_{H_d}^2$, and hence, 
the prediction does not rely on the particular model defined in Eqs.\,(\ref{eq:Kp}) and (\ref{eq:Sp}).
Therefore, 
in a class of models with $\bar{m}_{H_u}^2 \simeq \bar{m}_{H_d}^2$,
the electroweak fine-tuning predicts $\tan\b \simeq 1$  when
supersymmetry is broken at around the Planck scale.%
\footnote{
In a model with $\mu_H \ll m_{3/2}$ while $b = O(m_{3/2}^2)$,
the Higgsino can be a viable dark matter candidate when $\mu_H \gtrsim 10^8$\,GeV\,\cite{Feldstein:2013uha}.
}

\subsubsection*{Quartic coupling at the Planck scale}\vspace{-10pt}
Below the supersymmetry breaking scale at around the Planck scale, 
the Higgs sector consists of the light Higgs boson $h$ and its scalar potential is given by,
\begin{eqnarray}
V(h) = \frac{\l}{2}(h^\dagger h - v^2)^2 \ ,
\end{eqnarray}
where $v \simeq 174.1$\,GeV is achieved as a result of the fine-tuning of the quadratic terms as discussed above.
As a notable feature of the supersymmetric standard model, the Higgs quartic coupling $\lambda$ is given by the
 $SU(2)_L\times U(1)_Y$ gauge coupling constants, 
\begin{eqnarray}
\label{eq:lambda}
\lambda \simeq \frac{1}{4} \left(\frac{3}{5} g_1^2 +g_2^2\right) \cos^22\b\ ,
\end{eqnarray}
at the tree-level.
Thus, the prediction of $\tan \b \simeq 1$ from the electroweak fine-tuning 
results in the almost vanishing quartic coupling.

At the higher loop-level, the Higgs quartic coupling receives threshold corrections from the top squarks,
\begin{eqnarray}
{\mit \D}\l \simeq \frac{6y_t^4}{16\pi^2} \left(
\frac{X_t^2}{m_{\tilde t}^2} -\frac{1}{12} \frac{X_t^4}{m_{\tilde t}^4}\right)\ ,
\end{eqnarray}
where $X_t = A_t - \mu_H \cot\b$\,\cite{Okada:1990gg,Haber:1993an,Carena:1995wu}.
This contribution is, however, suppressed due to a small top Yukawa coupling at the Planck scale, $y_t \simeq 0.4$.

The Higgs quartic coupling also gets contributions from higher dimensional operators
which connect the supersymmetry breaking fields and the Higgs sector.
For example, higher dimensional operators
\begin{eqnarray}
\label{eq:HD}
{ \mit \D}K =   \frac{c}{M_{\rm PL}^4}|Z|^2| H_{u,d}|^4
\end{eqnarray}
lead to additional contributions to the quartic coupling,
\begin{eqnarray}
{\mit \D} \lambda = {\cal O}\left( c \frac{\L_{\rm SUSY}^4}{M_{\rm PL}^4} \right)\ .
\end{eqnarray}
These contributions are suppressed either when the supersymmetry
breaking sector is slightly separated from the Higgs sector, i.e. $c \ll 1$,
or when the supersymmetry breaking scale is somewhat smaller than the Planck scale.%
\footnote{For a related discussion, see also Ref.\,\cite{Okada:1990gg,Hall:2009nd}.}
It should be noted that the later possibility, e.g. $\Lambda_{\rm SUSY} \simeq 10^{17}$\,GeV,
does not affect the prediction of $\tan\b \simeq 1$.%
\footnote{
The predicted value of $\tan\b$ is significantly deviated from $1$
for a much lower supersymmetry breaking scale, $\L_{\rm SUSY }\ll 10^{17-18}$\,GeV,
where the renormalization group effects spoil the universality of the soft masses of the two Higgs doublets
even if the universal soft masses, $m_{H_u}^2 = m_{H_d}^2$, 
are realized at the mediation scale around the Planck scale.
}

\begin{figure}[t]
\begin{center}
  \includegraphics[width=0.75\linewidth]{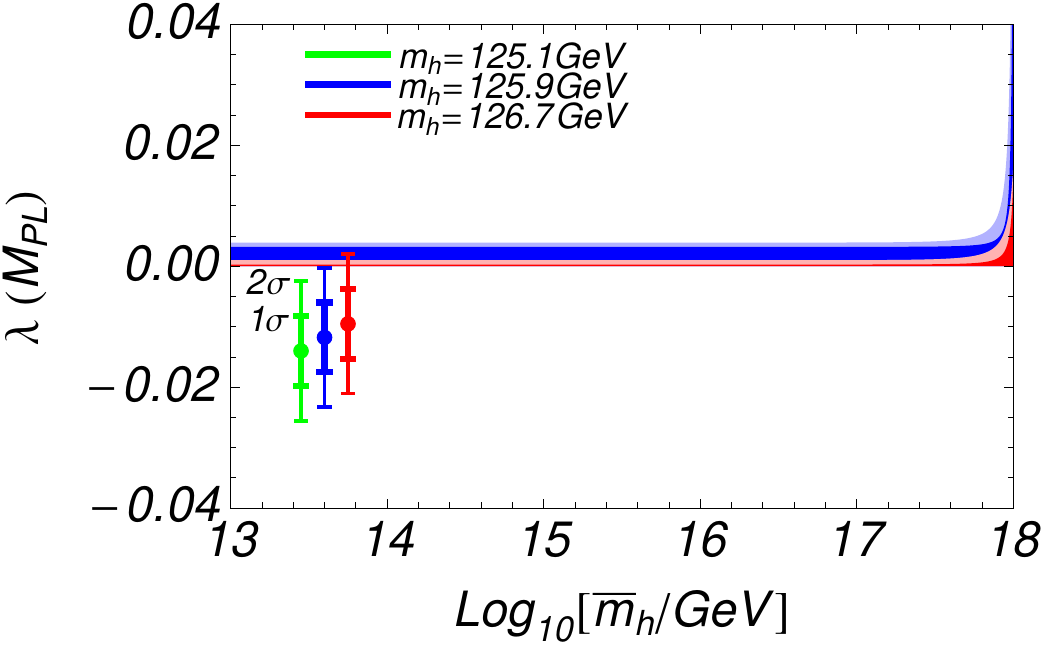}
\end{center}
\caption{\sl \small
The predicted Higgs quartic coupling at the Planck scale.
We allowed $X_t^2/m_{\tilde t}^2$ from $0$ to $10$.
The (light-)red shaded regions show the predicted quartic coupling for $X_t^2/m_{\tilde t}^2 = 0$
while allowing $m_{H_u}^2$ varying from $m_{H_d}^2$ by 10\% (20\%).
The (light)-blue shaded regions show the ones for $X_t^2/m_{\tilde t}^2 = 6$.
The values of the quartic coupling with error bars 
show the  Higgs quartic coupling extrapolated from the electroweak scale
for a given physical Higgs boson mass.
}
\label{fig:lambda}
\end{figure}
 
In Fig.\,\ref{fig:lambda},
we show the predicted quartic coupling at the Planck scale.
In our analysis, we assumed that  $X_t^2/m_{\tilde t}^2$ ranges between $0$ to $10$,
in which ${\mit\D} \lambda$ becomes maximal for $X_t^2/m_{\tilde t}^2\simeq 6$
for a given $\bar{m}_h$.%
\footnote{If we allow much larger value of  $X_t^2/m_{\tilde t}^2$, e.g. $X_t^2/m_{\tilde t}^2 \gtrsim 15$,
the predicted $\lambda$  takes a negative value.}
In the figure, the (light-)red shaded regions show the predicted quartic coupling for $X_t^2/m_{\tilde t}^2 = 0$
while allowing $m_{H_u}^2$ varying from $m_{H_d}^2$ by 10\% (20\%).
The (light)-blue shaded regions show the ones for $X_t^2/m_{\tilde t}^2 = 6$.
In the figure, we also show the Higgs quartic coupling extrapolated from the electroweak scale
assuming that there is no new physics below the Planck scale\,\cite{Degrassi:2012ry}.
The figure shows that the predicted quartic coupling is vanishingly small once 
the electroweak fine-tuning is required.
Therefore, we find that  the electroweak fine-tuning leads to a shallow Higgs potential 
in a class of models with $\bar{m}_{H_u}^2 \simeq \bar{m}_{H_d}^2$ when
supersymmetry is broken at around the Planck scale.%
\footnote{
It is also possible to provide $\lambda(M_{\rm PL}) \simeq 0$ if 
the $SU(2)_2\times U(1)_Y$ gaugino masses are dominated 
by the Dirac mass, which results in the vanishing $D$-term contributions
to the Higgs potential\,\cite{Fox:2002bu,Unwin:2012fj}. 
}

\subsubsection*{Discussions}\vspace{-10pt}
We have shown that the almost vanishing Higgs quartic coupling is predicted 
for $\bar{m}_{H_u}^2 \simeq \bar{m}_{H_d}^2$ with the Planck scale supersymmetry breaking.
It is an intriguing feature of this mechanism
that the shallowness of the Higgs potential 
is caused by the electroweak fine-tuning in the Higgs mass parameters.

Since the predicted quartic coupling is almost vanishing but is positive valued in 
most parameter space, the future precise measurements of the Higgs mass parameters
as well as the top Yukawa coupling and the strong coupling constants provide  
an important test of this mechanism.
At the ILC, for example, 
the Higgs mass can be determined with a precision of $~$30\,MeV
for the integrated luminosity ${\cal L} = 250$\,fb$^{-1}$\,\cite{Abe:2010aa,Li:2012taa}.
The uncertainties of the top Yukawa coupling  will be also 
reduced  by about one order of magnitude at the ILC\,\cite{Top}.
Improvements in lattice calculations could reduce the error of the strong coupling constant 
$\alpha_s$ down to $0.1$\%\cite{Campbell:2013qaa}.
With these improvements, it is possible to refute this mechanism if the
central value of the extrapolated Higgs quartic coupling at around the Planck scale 
is close to the current central values,
unless there is a small, but non-negligible contributions from Eq.\,(\ref{eq:HD}).

Finally, let us comment on a more ambitious interpretation of this mechanism.
In the simplest model we discussed in Eqs.\,(\ref{eq:Kp}) and (\ref{eq:Sp}),
the electroweak fine-tuning condition is nothing but the requirement of $c \simeq 1$.
In this case, the K\"ahler potential can be rewritten by $K = |H_u + H_d^\dagger|^2$,
and hence, the model has a shift symmetry, $H_{u,d} \to H_{u,d} + i \alpha$
with $\alpha$ being a real parameter.
This suggests that the prediction of $\tan\b \simeq 1$ can be related to the existence of
the shift symmetry.%
\footnote{
See also Refs.\,\cite{Hebecker:2012qp,Hebecker:2013lha} 
which discussed the connection between the prediction of $\tan\b \simeq 1$ 
and the shift symmetries as well as their realization in string theory.
}
In fact, the prediction of  $\tan\b \simeq 1$ is not altered even if we take 
a more generic K\"ahler potential as long as the shift symmetry is preserved, i.e. 
\begin{eqnarray}
K = K(H_u+H_d^\dagger, Z)\ .
\end{eqnarray}
Here, we do not need to assume that the couplings between the supersymmetry breaking sector and
the Higgs sector are  suppressed, since the above K\"ahler potential does not contribute to the scalar potential 
of the light Higgs boson, $h \simeq H_u - H_d^\dagger$.
It is notable that the shallowness of the Higgs potential can be interrelated to the 
shift symmetry of the Higgs sector
despite the fact that the shift symmetry is {\it explicitly broken}
by the gauge interactions which provide the leading contribution to the quartic coupling
in the supersymmetric standard model (Eq.\,(\ref{eq:lambda})).

It is also possible to extend this mechanism to more generic models in which 
the Higgs doublets emerge as Goldstone modes of approximate 
symmetries\,\cite{Kugo:1983ai,Kaplan:1983fs,Kaplan:1983sm,Inoue:1985cw}
such as models with $SU(3)/SU(2)\times U(1)$\,\cite{GY,Goto:1992hh, Hellerman:2013vxa}.
There, again, the prediction of $\tan\b\simeq 1$ is guaranteed by the non-linearly realized
symmetry by the Higgs doublets which non-trivially leads to the vanishing quartic coupling.

\subsubsection*{Note Added}
After this paper was posted to arXiv.org,  it came to the author's attention that Refs.\cite{Ibanez:2012zg,Ibanez:2013gf} 
have observed that the electroweak fine-tuning with the boundary condition with $m_{H_d}^2\simeq m_{H_d}^2$ 
at the intermediate to the scale of the unification leads to the appropriate Higgs boson mass, i.e. $m_H \simeq 126\pm 3$\,GeV.
These observations partially overlap with our arguments that the electroweak fine-tuning from the Planck scale supersymmetry
breaking leads to the flat Higgs potential at the Planck scale.
Their boundary conditions at the intermediate scale, however, may not be easily realized in a simple framework of supergravity 
due to non-negligible radiative corrections to the higgs boson masses.%
\footnote{We stress that the our setup does not require any specific mediation mechanisms of the supersymmetry
breaking effects other than  supergravity mediation.
}

\vspace{-10pt}
\section*{Acknowledgements} \vspace{-10pt}
The author would like to express his gratitude to L.~E.~Ibanez,  A.~Knochel, F.~Marchesano, A~Hebecker, T~Weigand,
for calling author's attention to important references.
This work is supported by the Grant-in-Aid for Scientific research from the
Ministry of Education, Science, Sports, and Culture (MEXT), Japan, No.\ 22244021 (TTY), 
No. 22244021 and No. 23740169 (S.M.), and No.\ 24740151 (MI).
It is also supported by the World Premier International Research Center Initiative (WPI Initiative), MEXT, Japan.

\end{document}